\documentclass[twocolumn,secnumarabic,amssymb, nobibnotes, aps, prd,superscriptaddress]{revtex4-2}
\usepackage{graphicx}
\usepackage{color}
\usepackage{amsmath}
\usepackage{gensymb}
\usepackage{tabularx}
\usepackage{multirow}

\usepackage[left,modulo]{lineno}
\usepackage{upgreek}

\begin{document} 
\title{Molecular shape controls dipolar correlations in polar liquids}

\author{Marceau H\'enot}
\email[Corresponding author: ]{marceau.henot@cea.fr}
\affiliation{SPEC, CEA, CNRS, Université Paris-Saclay, CEA Saclay Bat 772, 91191 Gif-sur-Yvette Cedex, France.}

\date{\today}
\begin{abstract}
The dielectric constant of polar liquids depends not only on the magnitude of the molecular dipole moment but also on short range orientational correlations between neighboring dipoles. How molecular structure determines these correlations, quantified by the Kirkwood correlation factor $g_\mathrm{K}$, remains unclear. Using molecular dynamics simulations of model polar liquids without hydrogen bonding, we show that molecular shape plays a central role. We interpret $g_\mathrm{K}$ as resulting from the combination of dipolar interactions and steric effects, which together favor either alignment or antialignment of neighboring molecules. Depending on molecular shape, $g_\mathrm{K}$ can either increases or decreases upon cooling, or change monotony. A minimal model of dipoles on a cubic lattice reproduces these three regimes and attributes the nonmonotonic case to a competition between the first layer of neighbors, which favors alignment, and the second layer, which favors antialignment.
\end{abstract}

\maketitle 

The static dielectric constant $\varepsilon$ of polar liquids sets the strength of electrostatic screening and thus governs phenomena such as ion solvation~\cite{israelachvili2011intermolecular}, miscibility of liquids~\cite{zhuang2021like} or chemical reactivity~\cite{reichardt2011solvents}. Predicting $\varepsilon$ from molecular properties, however, remains a long standing challenge in statistical physics: it requires understanding on how orientational correlations arise in a dense, disordered system of particles with anisotropic, long range interactions. Following the approach of Kirkwood and Fr\"ohlich~\cite{kirkwood1939dielectric, frohlich1948general, bottcher1973, kremer2002broadband}, the static dielectric constant can be written as: $\varepsilon - \varepsilon_\infty = \lambda F g_\mathrm{K}$, where $\varepsilon_\infty$ is the infinite frequency dielectric constant, $\lambda=\frac{\mu^2}{3\varepsilon_0 k_\mathrm{B}Tv}$ is the ratio of dipolar interaction to thermal energy (with $\mu$ the dipole moment, $v$ the molecular volume, $\varepsilon_0$ the vacuum permittivity, $k_\mathrm{B}$ the Boltzmann constant and $T$ the temperature), $F = \varepsilon(\varepsilon_\infty+2)^2/(3(2\varepsilon + \varepsilon_\infty))$ is the internal field factor~\cite{bottcher1973}, and $g_\mathrm{K}$ is the Kirkwood correlation factor accounting for orientational correlations between neighboring dipoles~\cite{kirkwood1939dielectric}. Predominantly parallel orientation is most often observed and correspond to $g_\mathrm{K}>1$ but some liquids display antiparallel correlations (\textit{i.e.} $g_\mathrm{K}<1$). The Kirkwood factor $g_\mathrm{K}$, estimated from measurements of $\varepsilon$, typically range from 0.5 to 5 for most systems~\cite{kremer2002broadband}. Since $\varepsilon$ scales almost linearly with $g_\mathrm{K}$ for strongly polar liquids, these variations translate into changes of the same relative magnitude in $\varepsilon$. This makes $g_\mathrm{K}$ the main source of difficulty in predicting, or even rationalizing, the values of $\varepsilon$ and their temperature dependence. 

In hydrogen bonded liquids, bond directionality strongly constrains the relative orientation of neighboring molecules~\cite{oster1943influence, bone2025chain} while bond strength stabilizes chainlike or network structures, whose reorganization can dominate the temperature dependence of $g_\mathrm{K}$~\cite{singh2012watching, bauer2013debye, henot2026molecular}. In the following, we focus on liquids not subject to hydrogen bonding, in which dipolar correlations are expected to arise from a different mechanism. This problem has been addressed in the past using empirical models based on large datasets~\cite{bouteloup2019predicting}, theoretical approaches~\cite{dejardin2018calculation, dejardin2022kinetic, dejardin2022temperature} or simulations of model molecules~\cite{johnson2010dielectric, johnson2015dielectric, koperwas2025can}, revealing the importance of the position of the polar group within a molecule. Yet, there is still no clear understanding on what controls the transition between alignment and anti-alignment or the reported variety of temperature dependence.

In this Letter, we used molecular dynamics simulations to study model polar molecules inspired by real systems. We systematically explored the effects of the dipole moment magnitude and of temperature on dipolar correlations for molecules with various shapes. We showed that, depending on the molecular shape, dipolar cross correlations could be always positive ($g_\mathrm{K}>1$), always negative  ($g_\mathrm{K}<1$), or undergo a transition from positive to negative as the strength of the dipolar interaction increases. We attributed this behavior to the interplay between dipolar interactions at short range and steric effects. We tested this hypothesis by developing a minimal lattice model of dipoles with effective interactions mimicking dipolar and steric interaction and we showed that it reproduces the entire phenomenology. This simple model helped us rationalizing the nonmonotonic temperature dependence as a competition between the first layer of neighbors promoting positive correlations and the second layer favoring anti-correlations.

\begin{figure}[htbp]
  \centering
  \includegraphics[width=\linewidth]{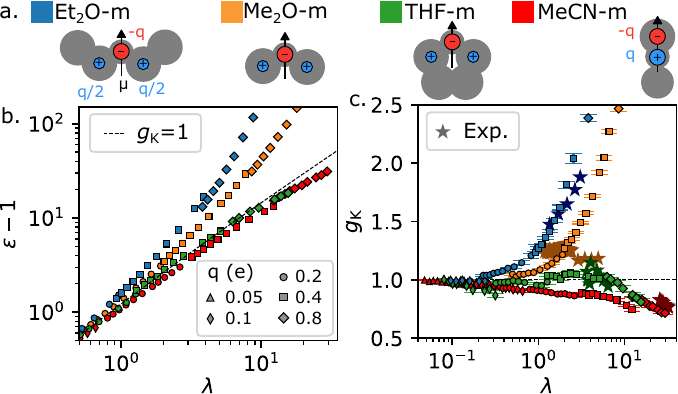}
 \caption{(a) Schematics of the model polar molecules considered in this study showing the distribution of partial charges and the resulting dipole. (b) Dielectric susceptibility $\varepsilon-1$ as a function of the dimensionless parameter $\lambda$ for the four model molecules (colors). (c) Kirkwood correlation factor as a function of $\lambda$. The star markers correspond  for each system to $g_\mathrm{K}$ calculated from experimental data~\cite{jacobs1944dielectric, gee1986dielectric,  blukis1963microwave, eltringham2007relative, engerholm1969ring, arivazhagan2011dielectric, nain2006densities, gadhi1995moment, stoppa2015dielectric}.}
  \label{fig1}
\end{figure}

To investigate how the magnitude of the dipole moment and the molecular shape together affect dipolar correlations, we selected four model systems derived from real molecules, shown in Fig.~\ref{fig1}a. Diethylether (Et$_2$O) and dimethylether (Me$_2$O) both have an elongated shape with a dipole moment perpendicular to their main axis. Tetrahydrofuran (THF) is a planar molecule with an inplane dipole moment. Acetonitrile (MeCN) is a rigid linear molecule whose dipole moment lies along its axis. For each molecule, we started from a united atom model and redistributed the partial charges so that the permanent dipole is carried by three atoms only, which reduces its sensitivity to conformational changes. The charges were then scaled so that the dipole moment $\mu$ ranges from 0.2~D to 4.5~D. This procedure yields four coarse-grained model systems Et$_2$O-m, Me$_2$O-m, THF-m and MeCN-m. For each one, equilibrium simulations were performed for several values of $\mu$ and several temperatures $T$ (see Appendix A). In this way, the dimensionless parameter $\lambda$ spans three orders of magnitude. The dielectric susceptibility $\varepsilon-1$ was obtained as a function of $\lambda$ from the dipolar fluctuations using the virtual cavity method~\cite{henot2025computing} to deal with the effect of electrostatic boundary conditions~\cite{zhang2016JPCL}, and the results are shown in Fig.~\ref{fig1}b. It is already visible that the molecular shape has a significant effect on $\varepsilon$ and that all systems deviate from the uncorrelated case ($g_\mathrm{K}=1$).

The Kirkwood correlation factor $g_\mathrm{K}$ of the infinite system was obtained as a function of $\lambda$, and the results are shown in Fig.~\ref{fig1}c and Fig.~\ref{figS1}. For all systems, $g_\mathrm{K}$ tends to 1 as $\lambda$ becomes small, as expected when thermal fluctuations dominate over dipolar interactions. Beyond this regime, the four systems behave very differently. For Et$_2$O-m and Me$_2$O-m, $g_\mathrm{K}(\lambda)$ increases monotonically, although with different rates. For THF-m, $g_\mathrm{K}$ stays close to or slightly above 1 before suddenly decreasing. Finally, for MeCN-m, $g_\mathrm{K}(\lambda)$ is always smaller than 1 and decreases continuously. The results of the simulations can be directly compared with experimental values of $g_\mathrm{K}(\lambda)$ obtained from measurements of $\varepsilon(T)$ for the four corresponding liquids~\cite{jacobs1944dielectric, gee1986dielectric,  blukis1963microwave, eltringham2007relative, engerholm1969ring, arivazhagan2011dielectric, nain2006densities, gadhi1995moment, stoppa2015dielectric} (see Appendix~B). These experimental results are shown as star markers in Fig.~\ref{fig1}c and are in very good agreement with the simulations for Et$_2$O, THF and MeCN, both regarding absolute value and temperature dependence. In the case of Me$_2$O, the experimental data are more scattered, and the simulations overestimates the temperature dependence. Such a level of agreement is nevertheless remarkable, given the simplicity of the model and the fact that the force fields were not optimized to reproduce the dielectric response~\cite{schmid2011definition}. Although the experimental data do not span a sufficiently wide range of $\lambda$ to reproduce the full $g_\mathrm{K}(\lambda)$ behavior observed in the simulations, they appear fully consistent with it.

%%%%%%%%%%%%%%%%%%%%%%%%%%%%%%%%%%%%%%%%
%%%%%%%%%%%%%%%%%%%%%%%%%%%%%%%%%%%%%%%%

\begin{figure}[htbp]
  \centering
  \includegraphics[width=\linewidth]{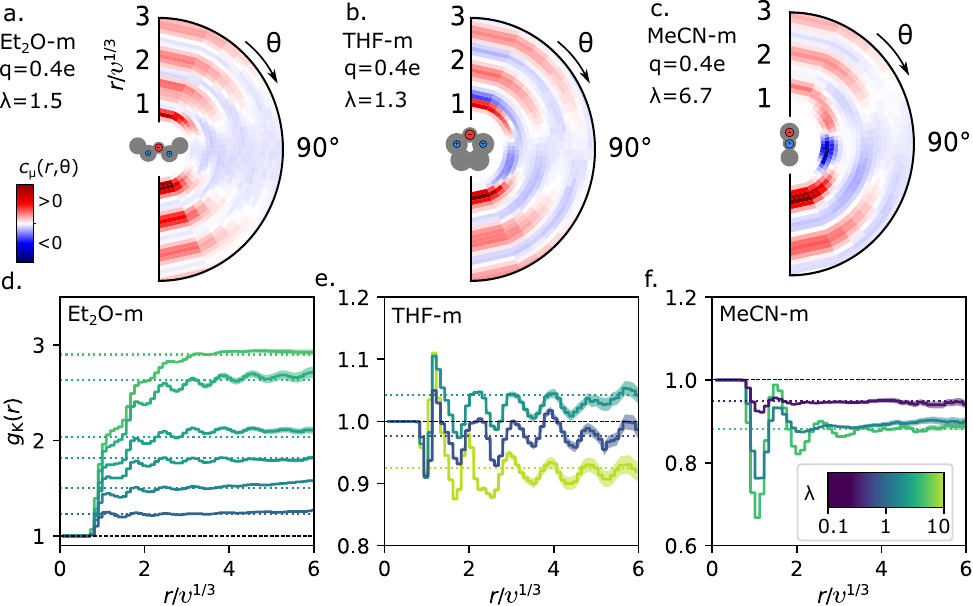}
 \caption{(a-c) Diagram of dipolar correlations as a function of distance and relative orientation for Et$_2$O-m (a), THF-m (b), and MeCN-m (c). (d-f) $r$-dependent Kirkwood correlation factor for Et$_2$O-m (d), THF-m (e) and MeCN-m (f) for different values of $\lambda$. Horizontal dotted lines correspond to $g_\mathrm{K}$.}
  \label{fig2}
\end{figure}

The Kirkwood factor $g_\mathrm{K}=1/N \sum_i\sum_j \vec{\mu}_i \cdot \vec{\mu}_j$ reflects the net dipolar correlations arising from a volume containing a large number $N$ of molecules. To understand its origin, it is useful to analyze how these correlations build up in space. To this end, we define, for a given reference dipole, the mean correlation $c_\mu(r, \theta)$ per unit distance and per unit of solid angle arising from neighboring dipoles located at a distance $r$ and at an angle $\theta$ with respect to the reference dipole (see Fig.~\ref{fig2}a-c). We can also define the $r$-dependent Kirkwood factor as the restriction of dipolar correlations to a sphere of radius $r$ around a reference dipole: $g_\mathrm{K}(r) = \int_0^{r}\iint_\Omega c_\mu \mathrm{d}\Omega\mathrm{d}\mathrm{r}$. In this framework, $g_\mathrm{K}$ corresponds to the plateau value of $g_\mathrm{K}(r)$ reached after a few molecular distance~\cite{zhang2016JPCL, henot2025computing}. Typical maps of $c_\mu(r, \theta)$ are shown for Et$_2$O-m, THF-m and MeCN-m in Fig.~\ref{fig2}a, b and c respectively. At large distances, the dipolar correlations are on average positive for molecules located above ($\theta=0$) and below ($\theta=\pi$) the reference dipole, and negative on its sides ($\theta=\pi/2$), as expected from the anisotropy of the interaction between dipoles. These contributions of opposite sign compensate each other after only a few molecular distance, as can be seen in Fig.~\ref{fig2}d, e and f from the plateau reached by $g_\mathrm{K}(r)$. For Et$_2$O, the distance at which the $g_\mathrm{K}$ is reached increases for large $\lambda$. This is associated with a steepening of $g_\mathrm{K}(\lambda)$ beyond the initial linear regime (see insets in Fig.~\ref{figS1}a and b) and might signal the proximity of a ferroelectric transition. Yet, such a transition is very rarely observed in real polar liquids (although a few examples have recently been reported~\cite{chen2020first}), and is out of the scope of the present work. 

Our first goal is to understand why Et$_2$O-m and MeCN-m favor respectively positive and negative net correlations. As shown by the maps of Fig.~\ref{fig2}a and c, the spatial distribution of the dipolar correlations at short range are very different for these two systems. For MeCN-m, the positive correlations at $\theta = 0$ and $\theta = \pi$ are compensated by strong negative correlations at $\theta=\pi/2$, which are almost absent for Et$_2$O-m. For a given value of $\lambda$, the difference between these two systems lies in their molecular shapes. For Et$_2$O-m, the lateral extension of the molecule prevents neighbors from occupying the positions where the dipolar interaction would most favor antiparallel alignment. In contrast, the linear shape of MeCN-m allows neighbors to occupy precisely these positions. This leads us to formulate the following hypothesis: the effect of molecular shape on the Kirkwood correlation factor, and hence on the dielectric permittivity, results from the combination of two effects: on the one hand, the dipolar interaction tends to align or anti-align neighboring molecules depending on their relative position. On the other hand, steric effects reduce the density of neighbors at specific orientations and thereby perturb the cancellation of dipolar correlations. This interpretation is consistent with previous simulations results of ellipsoidal polar particles, in which a dipole along the long axis favors local antiparallel alignment and keeps $g_\mathrm{K}$ close to unity~\cite{johnson2010dielectric, johnson2015dielectric}. Recently, a model molecule made of an elongated polar unit attached at one end of a larger apolar part was shown in simulations to lead to $g_\mathrm{K}<1$~\cite{koperwas2025can}. Our interpretation suggests a possible rationalization of this observation: the apolar part sterically obstruct the approach of neighbors along the dipole axis, where parallel alignment would be favored, while its very large dipole moment favor anti-alignment on its sides.

%%%%%%%%%%%%%%%%%%%%%%%%%%%%%%%%%%%%%%%%
%%%%%%%%%%%%%%%%%%%%%%%%%%%%%%%%%%%%%%%%

\begin{figure}[htbp]
	\centering
	\includegraphics[width=\linewidth]{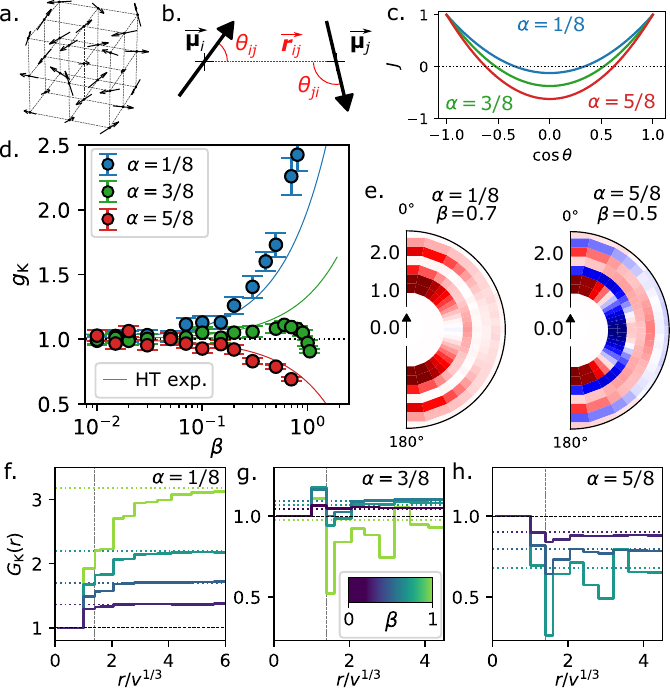}
	\caption{(a) Schematic of the dipoles on a lattice model. (b) Definition of the angle $\theta$ appearing in $\mathcal{H}$. (c) Interaction term $J(\theta)$ for the values of $\alpha$ considered in this study. (d) Kirkwood correlation factor as a function of $\beta=1/T$. Solid lines correspond to the linear high temperature expansion. (e) Diagram of dipolar correlations for two values of $\alpha$ (f,g,h) $r$-dependent Kirkwood factor for three values of $\alpha$ and several values of $\beta$ (color). Vertical dashed lines locate the range of interaction in $\mathcal{H}$.}
	\label{fig3}
\end{figure}

The simulations of polar molecules presented above provide a simplified description of reality: electronic polarizability and dispersion interactions are absent, molecular shapes are coarse-grained, and dipole moments are rigid. Yet, the Hamiltonian of the system still retains a fair degree of complexity: the intermolecular interactions include dipolar interactions acting at long range, while steric effects are set by the detailed molecular shape and by the Lennard Jones potentials. And it visible in Fig.~\ref{fig1}c and \ref{figS1} that for all systems studied here, the collapse of $g_\mathrm{K}$ with $\lambda$ is not perfect when $\mu$ is varied. This is likely due to the fact that the molecular shape itself can affect orientational correlations in a way that does not scale with $\lambda$. To test the hypothesis formulated above, we now consider a minimal model retaining only the main physical ingredients of this hypothesis. It consists of $N$ dipoles $\vec{\mu}$ of unit magnitude, free to rotate but whose positions are fixed on a simple cubic lattice as shown in Fig.~\ref{fig3}a. The Hamiltonian of the system is:
\begin{align}
	\mathcal{H} = -\frac{1}{2}\sum_i\sum_{j\in v_i} J(\theta_{ij})\vec{\mu}_i\cdot \vec{\mu}_j \label{eq_H_MC} \\
	J(\theta) = 1-\alpha + (1 + \alpha)\cos 2\theta
\end{align}
The dependence of $J$ on $\theta$ mimics the dipolar interaction by favoring positive correlations at $\theta=0$ and $\theta=\pi$ and negative correlations at $\theta=\pi/2$ but the range is limited to the first $s=6$ neighbors. The parameter $\alpha$ tunes how favorable anticorrelations are at the lateral positions. This mimics steric constraints and acts as an effective description of the molecular shape. We performed Monte Carlo simulations (see Appendix~C) of the Hamiltonian of Eq.~\ref{eq_H_MC} for three positive values of $\alpha$ with the corresponding profiles of $J(\theta)$ plotted in Fig.~\ref{fig3}c. From these simulations, we extracted the Kirkwood correlation factor, shown as a function of the inverse temperature $\beta$ in Fig.~\ref{fig3}d. Since $J(\theta)$ is always of order unity, $\beta$ plays here the same role as $\lambda$ above: it measures the ratio between the dipolar interaction energy and the thermal energy.

%%%%%%%%%%%%%%%%%%%%%%%%%%%%%%%%%%%%%%%%
%%%%%%%%%%%%%%%%%%%%%%%%%%%%%%%%%%%%%%%%

As shown in Fig.~\ref{fig3}d, by modulating the value of $\alpha$, the lattice model is able to qualitatively reproduce the three regimes observed for model polar liquids. For $\alpha = 1/8$, $g_\mathrm{K}$ follows a monotonous increase on the considered range of $\beta$, as observed for Et$_2$O-m and Me$_2$O-m. For $\alpha=3/8$, the behavior is very similar to that of THF-m: $g_\mathrm{K}$ remains close to 1 before quickly decreasing at finite $\beta$. Finally for $\alpha=5/8$, $g_\mathrm{K}$ is monotonically decreasing and is always smaller than 1 as in MeCN-m. The comparison can be pushed further: the spatial distribution of dipolar correlations (see Fig.~\ref{fig3}e) and the evolution of the $r$-dependent Kirkwood factor (see Fig.~\ref{fig3}f, g and h) are very similar to that of Et$_2$O-m and MeCN-m respectively (see Fig.~\ref{fig2}). The Hamiltonian of Eq.~\ref{eq_H_MC} can be expanded at high temperature, leading to $g_\mathrm{K}(\beta) \approx 1+sJ_\mathrm{eff}\beta/3$ where $J_\mathrm{eff}=\langle J(\theta)\rangle = 2/3(1-2\alpha)$ is the effective interaction term. This linear behavior is shown with solid lines in Fig.~\ref{fig3}d and, as expected, describe well the evolution of $g_\mathrm{K}(\beta)$ at high temperature. 

%%%%%%%%%%%%%%%%%%%%%%%%%%%%%%%%%%%%%%%%
\begin{figure}[htbp]
	\centering
	\includegraphics[width=\linewidth]{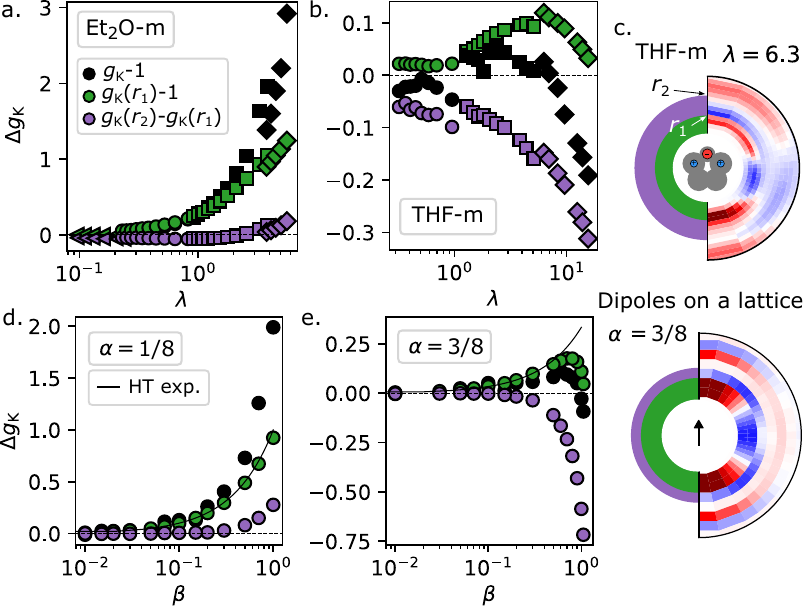}
	\caption{Decomposition of the dipolar cross correlation originating from the first layer of neighbors (green), the second layer (purple), or all neighbors (black). The decomposition of the first coordination shell into two layers is illustrated in (c). Results for polar liquids are shown for Et$_2$O-m (a) and THF-m (b) with same legend as in Fig.~\ref{fig1} for the markers. Results for the lattice model are shown for $\alpha=1/8$ (d) and $\alpha=3/8$ (e).}
	\label{fig4}
\end{figure}
%%%%%%%%%%%%%%%%%%%%%%%%%%%%%%%%%%%%%%%%

This simple model confirms the hypothesis that the combination of dipolar like interactions and steric effects at short range is sufficient to reproduce the phenomenology observed for the polar liquids. The two limiting cases, in which the dipolar cross correlation is either always positive ($g_\mathrm{K}(\beta) > 1$ for $\alpha = 1/8$) or always negative ($g_\mathrm{K}(\beta) < 1$ for $\alpha = 5/8$), are naturally explained by the fact that parallel or antiparallel orientations, respectively, are always favored at short range ($J_\mathrm{eff}>0$ or $J_\mathrm{eff}<0$). In both these cases, the deviations from the high temperature expansion only reinforce this effect. For $\alpha = 1/8$, they clearly arise from the contribution of more distant neighbors (see Fig.~\ref{fig3}f) as observed for Et$_2$O and Me$_2$O. In contrast, the change of sign of the dipolar cross correlation at finite temperature, observed for $\alpha = 3/8$ and THF-m, is much less intuitive \textit{a priori}. The lattice model can be used to elucidate the physical mechanism behind this nonmonotonic behavior of $g_\mathrm{K}(\beta)$.

The $r$-dependent Kirkwood factor $g_\mathrm{K}(r)$ shown for $\alpha=3/8$ in Fig.~\ref{fig3}g, reveals that the first short range feature, corresponding to the 6 nearest neighbors, is consistent with the positive sign of $J_\mathrm{eff}$, whereas the next layer correlates in the opposite direction. To further characterize this effect, we plot in Figs.~\ref{fig4}d and e the evolution with $\beta$ of three quantities: the net cross correlation arising from the 6 nearest neighbors, $g_\mathrm{K}(r_1) - 1$; the net contribution of the next layer, $g_\mathrm{K}(r_2) - g_\mathrm{K}(r_1)$; and the total net cross correlation, $g_\mathrm{K}-1$. It is important to note that only the first layer is directly coupled to the reference dipole through the Hamiltonian, while the correlation of the next layer is mediated by the first one. As shown in Figs.~\ref{fig3}e-right and \ref{fig4}c-bottom, this second layer tends to anticorrelate along $\theta = \pi/4$. This corresponds to a favorable orientation with respect to the first layer: a negative correlation with the dipole at $\theta = 0$ and a positive correlation with the one at $\theta = \pi/2$. This suggests the following physical picture: for certain values of $\alpha$, the first layer can exhibit a net positive correlation while promoting sufficient anticorrelation in the lateral direction for the next layer to display a systematic tendency toward net anticorrelation. This is the case for $\alpha=3/8$, where the nearest neighbors dominate at high temperature but are progressively overtaken by the next layer, leading to a nonmonotonic behavior of $g_\mathrm{K}(\beta)$ (see Fig.~\ref{fig4}e). At sufficiently low temperature, even the contribution of the first layer starts to deviate from the linear expansion, likely because of the influence of more distant neighbors, which contribute increasingly to the net correlation until $g_\mathrm{K}$ ultimately becomes smaller than 1. In contrast, for $\alpha=1/8$ (see Fig.~\ref{fig3}f and \ref{fig4}d) and $\alpha=5/8$ (see Fig.~\ref{fig3}h), the second layer only reinforces the first one, so that $g_\mathrm{K}(\beta)$ remains monotonic. This explanation can be tested for the simulations of polar liquids for which we decompose the first coordination shell into two layers as shown in Fig.~\ref{fig4}c for THF-m with more details given in appendix D. The correlations resulting from the first two layers are shown for Et$_2$O-m and THF-m in Fig.~\ref{fig4}a and b and show a behavior strikingly similar to that of the $\alpha=1/8$ and $\alpha=3/8$ cases respectively, reinforcing the idea that the origin of the temperature dependence of $g_\mathrm{K}$ are the same. These results can be compared to the theory developed by Déjardin for the Kirkwood factor of interacting dipoles with spherical or needle shape~\cite{dejardin2022kinetic}. In its simplest form including only dipolar interaction, it predicts the possibility of the two monotonous regimes of $g_\mathrm{K}>1$ and $g_\mathrm{K}<1$~\cite{dejardin2018calculation}. By including the effect of induction and dispersion interaction, it allows for a regime of nonmonotonic $g_\mathrm{K}(\lambda)$ similar of the one we observed for THF-m or the lattice model with $\alpha=3/8$. Although electronic polarizability could in principle affect $g_\mathrm{K}$ beyond the reaction field effect, the fact that the present results can be qualitatively compare with the theory despite the difference in physical ingredient may results from the fact that the term accounting for these interaction could effectively be used to account for the effect of the molecular shape.

To conclude, by combining simulations of model polar liquids not subject to hydrogen bonding with realistic shapes with a simplified model of dipoles on a lattice, we evidenced three regime of dipolar correlations when varying the ratio between dipolar interaction and thermal energy $\lambda$. First, a regime favoring alignment, corresponding to a $g_\mathrm{K}>1$ and increasing upon cooling, relevant for a majority of polar liquids. Second, a regime favoring anti-alignment with $g_\mathrm{K}<1$ and decreasing upon cooling, relevant for liquids exhibiting a relatively low dielectric constant given their large dipole moment, such as acetonitrile. Finally, a regime exhibiting a transition from favored alignment to anti-alignment when $\lambda$ is increased. This can explain values of $g_\mathrm{K}$ close or above unity but decreasing upon cooling, as observed in THF or in propylene carbonate~\cite{simeral1970dielectric, sengwa2017characterization}. We attribute these different regimes to a combination of dipolar correlations favoring specific orientation of close neighbors and steric effect leading to either a net alignment or anti-alignment of the first layer. The next layer can show a tendency to anti-alignment which, if strong enough, can become dominant at low temperature and lead to a non-monotonous $g_\mathrm{K}(\lambda)$. A natural perspective of this work is the question of the dynamics of the dipolar correlations. Indeed, these correlations not only affect the static dielectric constant $\varepsilon$ but also the shape of dielectric spectra to a significant extent~\cite{koperwas2022computational, pabst2021generic, henot2023orientational, matyushov2023single, bohmer2024spectral, bohmer2024dipolar, henot2025emergence, koperwas2025can}. Simulations of model polar liquids in which dipolar correlations can be modulated will be useful for investigating their consequences for the molecular dynamics probed by dielectric spectroscopy.

\textbf{Acknowledgments.} The author thanks F. Ladieu and P.M. Déjardin for fruitful discussions and for providing feedback on the manuscript.

\textbf{Data availability.} The data that support the findings of this article are not publicly available. The data are available from the author upon reasonable request.

\bibliographystyle{apsrev4-2}
%\bibliography{biblio}

\begin{thebibliography}{45}%
	\makeatletter
	\providecommand \@ifxundefined [1]{%
		\@ifx{#1\undefined}
	}%
	\providecommand \@ifnum [1]{%
		\ifnum #1\expandafter \@firstoftwo
		\else \expandafter \@secondoftwo
		\fi
	}%
	\providecommand \@ifx [1]{%
		\ifx #1\expandafter \@firstoftwo
		\else \expandafter \@secondoftwo
		\fi
	}%
	\providecommand \natexlab [1]{#1}%
	\providecommand \enquote  [1]{``#1''}%
	\providecommand \bibnamefont  [1]{#1}%
	\providecommand \bibfnamefont [1]{#1}%
	\providecommand \citenamefont [1]{#1}%
	\providecommand \href@noop [0]{\@secondoftwo}%
	\providecommand \href [0]{\begingroup \@sanitize@url \@href}%
	\providecommand \@href[1]{\@@startlink{#1}\@@href}%
	\providecommand \@@href[1]{\endgroup#1\@@endlink}%
	\providecommand \@sanitize@url [0]{\catcode `\\12\catcode `\$12\catcode
		`\&12\catcode `\#12\catcode `\^12\catcode `\_12\catcode `\%12\relax}%
	\providecommand \@@startlink[1]{}%
	\providecommand \@@endlink[0]{}%
	\providecommand \url  [0]{\begingroup\@sanitize@url \@url }%
	\providecommand \@url [1]{\endgroup\@href {#1}{\urlprefix }}%
	\providecommand \urlprefix  [0]{URL }%
	\providecommand \Eprint [0]{\href }%
	\providecommand \doibase [0]{https://doi.org/}%
	\providecommand \selectlanguage [0]{\@gobble}%
	\providecommand \bibinfo  [0]{\@secondoftwo}%
	\providecommand \bibfield  [0]{\@secondoftwo}%
	\providecommand \translation [1]{[#1]}%
	\providecommand \BibitemOpen [0]{}%
	\providecommand \bibitemStop [0]{}%
	\providecommand \bibitemNoStop [0]{.\EOS\space}%
	\providecommand \EOS [0]{\spacefactor3000\relax}%
	\providecommand \BibitemShut  [1]{\csname bibitem#1\endcsname}%
	\let\auto@bib@innerbib\@empty
	%</preamble>
	\bibitem [{\citenamefont
		{Israelachvili}(2011)}]{israelachvili2011intermolecular}%
	\BibitemOpen
	\bibfield  {author} {\bibinfo {author} {\bibfnamefont {J.~N.}\ \bibnamefont
			{Israelachvili}},\ }\href@noop {} {\emph {\bibinfo {title} {Intermolecular
				and surface forces}}}\ (\bibinfo  {publisher} {Academic press},\ \bibinfo
	{year} {2011})\BibitemShut {NoStop}%
	\bibitem [{\citenamefont {Zhuang}\ \emph {et~al.}(2021)\citenamefont {Zhuang},
		\citenamefont {Ramanauskaite}, \citenamefont {Koa},\ and\ \citenamefont
		{Wang}}]{zhuang2021like}%
	\BibitemOpen
	\bibfield  {author} {\bibinfo {author} {\bibfnamefont {B.}~\bibnamefont
			{Zhuang}}, \bibinfo {author} {\bibfnamefont {G.}~\bibnamefont
			{Ramanauskaite}}, \bibinfo {author} {\bibfnamefont {Z.~Y.}\ \bibnamefont
			{Koa}},\ and\ \bibinfo {author} {\bibfnamefont {Z.-G.}\ \bibnamefont
			{Wang}},\ }\href@noop {} {\bibfield  {journal} {\bibinfo  {journal} {Science
				advances}\ }\textbf {\bibinfo {volume} {7}},\ \bibinfo {pages} {eabe7275}
		(\bibinfo {year} {2021})}\BibitemShut {NoStop}%
	\bibitem [{\citenamefont {Reichardt}\ and\ \citenamefont
		{Welton}(2011)}]{reichardt2011solvents}%
	\BibitemOpen
	\bibfield  {author} {\bibinfo {author} {\bibfnamefont {C.}~\bibnamefont
			{Reichardt}}\ and\ \bibinfo {author} {\bibfnamefont {T.}~\bibnamefont
			{Welton}},\ }\href@noop {} {\emph {\bibinfo {title} {Solvents and solvent
				effects in organic chemistry}}}\ (\bibinfo  {publisher} {John Wiley \&
		Sons},\ \bibinfo {year} {2011})\BibitemShut {NoStop}%
	\bibitem [{\citenamefont {Kirkwood}(1939)}]{kirkwood1939dielectric}%
	\BibitemOpen
	\bibfield  {author} {\bibinfo {author} {\bibfnamefont {J.}~\bibnamefont
			{Kirkwood}},\ }\href {https://doi.org/10.1063/1.1750343} {\bibfield
		{journal} {\bibinfo  {journal} {The Journal of Chemical Physics}\ }\textbf
		{\bibinfo {volume} {7}},\ \bibinfo {pages} {911} (\bibinfo {year} {1939})},\
	\Eprint {https://arxiv.org/abs/https://doi.org/10.1063/1.1750343}
	{https://doi.org/10.1063/1.1750343} \BibitemShut {NoStop}%
	\bibitem [{\citenamefont {Fr{\"o}hlich}(1948)}]{frohlich1948general}%
	\BibitemOpen
	\bibfield  {author} {\bibinfo {author} {\bibfnamefont {H.}~\bibnamefont
			{Fr{\"o}hlich}},\ }\href@noop {} {\bibfield  {journal} {\bibinfo  {journal}
			{Transactions of the Faraday Society}\ }\textbf {\bibinfo {volume} {44}},\
		\bibinfo {pages} {238} (\bibinfo {year} {1948})}\BibitemShut {NoStop}%
	\bibitem [{\citenamefont {Böttcher}(1973)}]{bottcher1973}%
	\BibitemOpen
	\bibfield  {author} {\bibinfo {author} {\bibfnamefont {C.}~\bibnamefont
			{Böttcher}},\ }\href@noop {} {\emph {\bibinfo {title} {Theory of electric
				polarization: Dielectrics in static fields}}},\ \bibinfo {number} {vol. 1}\
	(\bibinfo  {publisher} {Elsevier},\ \bibinfo {year} {1973})\BibitemShut
	{NoStop}%
	\bibitem [{\citenamefont {Kremer}\ and\ \citenamefont
		{Sch{\"o}nhals}(2002)}]{kremer2002broadband}%
	\BibitemOpen
	\bibfield  {author} {\bibinfo {author} {\bibfnamefont {F.}~\bibnamefont
			{Kremer}}\ and\ \bibinfo {author} {\bibfnamefont {A.}~\bibnamefont
			{Sch{\"o}nhals}},\ }\href@noop {} {\emph {\bibinfo {title} {Broadband
				dielectric spectroscopy}}}\ (\bibinfo  {publisher} {Springer Science \&
		Business Media},\ \bibinfo {year} {2002})\BibitemShut {NoStop}%
	\bibitem [{\citenamefont {Oster}\ and\ \citenamefont
		{Kirkwood}(1943)}]{oster1943influence}%
	\BibitemOpen
	\bibfield  {author} {\bibinfo {author} {\bibfnamefont {G.}~\bibnamefont
			{Oster}}\ and\ \bibinfo {author} {\bibfnamefont {J.~G.}\ \bibnamefont
			{Kirkwood}},\ }\href@noop {} {\bibfield  {journal} {\bibinfo  {journal}
			{Journal of Chemical Physics}\ }\textbf {\bibinfo {volume} {11}},\ \bibinfo
		{pages} {175} (\bibinfo {year} {1943})}\BibitemShut {NoStop}%
	\bibitem [{\citenamefont {Bone}\ \emph {et~al.}(2025)\citenamefont {Bone},
		\citenamefont {Chung}, \citenamefont {Ponder},\ and\ \citenamefont
		{Schwarz}}]{bone2025chain}%
	\BibitemOpen
	\bibfield  {author} {\bibinfo {author} {\bibfnamefont {R.~A.}\ \bibnamefont
			{Bone}}, \bibinfo {author} {\bibfnamefont {M.~K.}\ \bibnamefont {Chung}},
		\bibinfo {author} {\bibfnamefont {J.~W.}\ \bibnamefont {Ponder}},\ and\
		\bibinfo {author} {\bibfnamefont {K.}~\bibnamefont {Schwarz}},\ }\href@noop
	{} {\bibfield  {journal} {\bibinfo  {journal} {The Journal of Physical
				chemistry. B}\ }\textbf {\bibinfo {volume} {129}},\ \bibinfo {pages} {8946}
		(\bibinfo {year} {2025})}\BibitemShut {NoStop}%
	\bibitem [{\citenamefont {Singh}\ and\ \citenamefont
		{Richert}(2012)}]{singh2012watching}%
	\BibitemOpen
	\bibfield  {author} {\bibinfo {author} {\bibfnamefont {L.~P.}\ \bibnamefont
			{Singh}}\ and\ \bibinfo {author} {\bibfnamefont {R.}~\bibnamefont
			{Richert}},\ }\href@noop {} {\bibfield  {journal} {\bibinfo  {journal}
			{Physical Review Letters}\ }\textbf {\bibinfo {volume} {109}},\ \bibinfo
		{pages} {167802} (\bibinfo {year} {2012})}\BibitemShut {NoStop}%
	\bibitem [{\citenamefont {Bauer}\ \emph {et~al.}(2013)\citenamefont {Bauer},
		\citenamefont {Burlafinger}, \citenamefont {Gainaru}, \citenamefont
		{Lunkenheimer}, \citenamefont {Hiller}, \citenamefont {Loidl},\ and\
		\citenamefont {B{\"o}hmer}}]{bauer2013debye}%
	\BibitemOpen
	\bibfield  {author} {\bibinfo {author} {\bibfnamefont {S.}~\bibnamefont
			{Bauer}}, \bibinfo {author} {\bibfnamefont {K.}~\bibnamefont {Burlafinger}},
		\bibinfo {author} {\bibfnamefont {C.}~\bibnamefont {Gainaru}}, \bibinfo
		{author} {\bibfnamefont {P.}~\bibnamefont {Lunkenheimer}}, \bibinfo {author}
		{\bibfnamefont {W.}~\bibnamefont {Hiller}}, \bibinfo {author} {\bibfnamefont
			{A.}~\bibnamefont {Loidl}},\ and\ \bibinfo {author} {\bibfnamefont
			{R.}~\bibnamefont {B{\"o}hmer}},\ }\href@noop {} {\bibfield  {journal}
		{\bibinfo  {journal} {The Journal of chemical physics}\ }\textbf {\bibinfo
			{volume} {138}},\ \bibinfo {pages} {094505} (\bibinfo {year}
		{2013})}\BibitemShut {NoStop}%
	\bibitem [{\citenamefont {H{\'e}not}\ and\ \citenamefont
		{Gabriel}(2026)}]{henot2026molecular}%
	\BibitemOpen
	\bibfield  {author} {\bibinfo {author} {\bibfnamefont {M.}~\bibnamefont
			{H{\'e}not}}\ and\ \bibinfo {author} {\bibfnamefont {J.~P.}\ \bibnamefont
			{Gabriel}},\ }\href@noop {} {\bibfield  {journal} {\bibinfo  {journal}
			{Physical Review E}\ }\textbf {\bibinfo {volume} {113}},\ \bibinfo {pages}
		{L043403} (\bibinfo {year} {2026})}\BibitemShut {NoStop}%
	\bibitem [{\citenamefont {Bouteloup}\ and\ \citenamefont
		{Mathieu}(2019)}]{bouteloup2019predicting}%
	\BibitemOpen
	\bibfield  {author} {\bibinfo {author} {\bibfnamefont {R.}~\bibnamefont
			{Bouteloup}}\ and\ \bibinfo {author} {\bibfnamefont {D.}~\bibnamefont
			{Mathieu}},\ }\href@noop {} {\bibfield  {journal} {\bibinfo  {journal}
			{Physical Chemistry Chemical Physics}\ }\textbf {\bibinfo {volume} {21}},\
		\bibinfo {pages} {11043} (\bibinfo {year} {2019})}\BibitemShut {NoStop}%
	\bibitem [{\citenamefont {D{\'e}jardin}\ \emph {et~al.}(2018)\citenamefont
		{D{\'e}jardin}, \citenamefont {Cornaton}, \citenamefont {Ghesqui{\`e}re},
		\citenamefont {Caliot},\ and\ \citenamefont
		{Brouzet}}]{dejardin2018calculation}%
	\BibitemOpen
	\bibfield  {author} {\bibinfo {author} {\bibfnamefont {P.-M.}\ \bibnamefont
			{D{\'e}jardin}}, \bibinfo {author} {\bibfnamefont {Y.}~\bibnamefont
			{Cornaton}}, \bibinfo {author} {\bibfnamefont {P.}~\bibnamefont
			{Ghesqui{\`e}re}}, \bibinfo {author} {\bibfnamefont {C.}~\bibnamefont
			{Caliot}},\ and\ \bibinfo {author} {\bibfnamefont {R.}~\bibnamefont
			{Brouzet}},\ }\href@noop {} {\bibfield  {journal} {\bibinfo  {journal} {The
				Journal of Chemical Physics}\ }\textbf {\bibinfo {volume} {148}},\ \bibinfo
		{pages} {044504} (\bibinfo {year} {2018})}\BibitemShut {NoStop}%
	\bibitem [{\citenamefont {D{\'e}jardin}(2022)}]{dejardin2022kinetic}%
	\BibitemOpen
	\bibfield  {author} {\bibinfo {author} {\bibfnamefont {P.-M.}\ \bibnamefont
			{D{\'e}jardin}},\ }\href@noop {} {\bibfield  {journal} {\bibinfo  {journal}
			{Physical Review E}\ }\textbf {\bibinfo {volume} {105}},\ \bibinfo {pages}
		{024109} (\bibinfo {year} {2022})}\BibitemShut {NoStop}%
	\bibitem [{\citenamefont {D{\'e}jardin}\ \emph {et~al.}(2022)\citenamefont
		{D{\'e}jardin}, \citenamefont {Pabst}, \citenamefont {Cornaton},
		\citenamefont {Helbling},\ and\ \citenamefont
		{Blochowicz}}]{dejardin2022temperature}%
	\BibitemOpen
	\bibfield  {author} {\bibinfo {author} {\bibfnamefont {P.-M.}\ \bibnamefont
			{D{\'e}jardin}}, \bibinfo {author} {\bibfnamefont {F.}~\bibnamefont {Pabst}},
		\bibinfo {author} {\bibfnamefont {Y.}~\bibnamefont {Cornaton}}, \bibinfo
		{author} {\bibfnamefont {A.}~\bibnamefont {Helbling}},\ and\ \bibinfo
		{author} {\bibfnamefont {T.}~\bibnamefont {Blochowicz}},\ }\href@noop {}
	{\bibfield  {journal} {\bibinfo  {journal} {Physical Review E}\ }\textbf
		{\bibinfo {volume} {105}},\ \bibinfo {pages} {024108} (\bibinfo {year}
		{2022})}\BibitemShut {NoStop}%
	\bibitem [{\citenamefont {Johnson}\ \emph {et~al.}(2010)\citenamefont
		{Johnson}, \citenamefont {Barnes}, \citenamefont {Draxler}, \citenamefont
		{Eichinger},\ and\ \citenamefont {Robinson}}]{johnson2010dielectric}%
	\BibitemOpen
	\bibfield  {author} {\bibinfo {author} {\bibfnamefont {L.~E.}\ \bibnamefont
			{Johnson}}, \bibinfo {author} {\bibfnamefont {R.}~\bibnamefont {Barnes}},
		\bibinfo {author} {\bibfnamefont {T.~W.}\ \bibnamefont {Draxler}}, \bibinfo
		{author} {\bibfnamefont {B.~E.}\ \bibnamefont {Eichinger}},\ and\ \bibinfo
		{author} {\bibfnamefont {B.~H.}\ \bibnamefont {Robinson}},\ }\href@noop {}
	{\bibfield  {journal} {\bibinfo  {journal} {The Journal of Physical Chemistry
				B}\ }\textbf {\bibinfo {volume} {114}},\ \bibinfo {pages} {8431} (\bibinfo
		{year} {2010})}\BibitemShut {NoStop}%
	\bibitem [{\citenamefont {Johnson}\ \emph {et~al.}(2015)\citenamefont
		{Johnson}, \citenamefont {Benight}, \citenamefont {Barnes},\ and\
		\citenamefont {Robinson}}]{johnson2015dielectric}%
	\BibitemOpen
	\bibfield  {author} {\bibinfo {author} {\bibfnamefont {L.~E.}\ \bibnamefont
			{Johnson}}, \bibinfo {author} {\bibfnamefont {S.~J.}\ \bibnamefont
			{Benight}}, \bibinfo {author} {\bibfnamefont {R.}~\bibnamefont {Barnes}},\
		and\ \bibinfo {author} {\bibfnamefont {B.~H.}\ \bibnamefont {Robinson}},\
	}\href@noop {} {\bibfield  {journal} {\bibinfo  {journal} {The Journal of
				Physical Chemistry B}\ }\textbf {\bibinfo {volume} {119}},\ \bibinfo {pages}
		{5240} (\bibinfo {year} {2015})}\BibitemShut {NoStop}%
	\bibitem [{\citenamefont {Koperwas}\ and\ \citenamefont
		{Paluch}(2025)}]{koperwas2025can}%
	\BibitemOpen
	\bibfield  {author} {\bibinfo {author} {\bibfnamefont {K.}~\bibnamefont
			{Koperwas}}\ and\ \bibinfo {author} {\bibfnamefont {M.}~\bibnamefont
			{Paluch}},\ }\href@noop {} {\bibfield  {journal} {\bibinfo  {journal} {New
				Journal of Physics}\ }\textbf {\bibinfo {volume} {27}},\ \bibinfo {pages}
		{093504} (\bibinfo {year} {2025})}\BibitemShut {NoStop}%
	\bibitem [{\citenamefont {H{\'e}not}(2025)}]{henot2025computing}%
	\BibitemOpen
	\bibfield  {author} {\bibinfo {author} {\bibfnamefont {M.}~\bibnamefont
			{H{\'e}not}},\ }\href@noop {} {\bibfield  {journal} {\bibinfo  {journal} {The
				Journal of chemical physics}\ }\textbf {\bibinfo {volume} {163}},\ \bibinfo
		{pages} {124501} (\bibinfo {year} {2025})}\BibitemShut {NoStop}%
	\bibitem [{\citenamefont {Zhang}\ \emph {et~al.}(2016)\citenamefont {Zhang},
		\citenamefont {Hutter},\ and\ \citenamefont {Sprik}}]{zhang2016JPCL}%
	\BibitemOpen
	\bibfield  {author} {\bibinfo {author} {\bibfnamefont {C.}~\bibnamefont
			{Zhang}}, \bibinfo {author} {\bibfnamefont {J.}~\bibnamefont {Hutter}},\ and\
		\bibinfo {author} {\bibfnamefont {M.}~\bibnamefont {Sprik}},\ }\href@noop {}
	{\bibfield  {journal} {\bibinfo  {journal} {The journal of physical chemistry
				letters}\ }\textbf {\bibinfo {volume} {7}},\ \bibinfo {pages} {2696}
		(\bibinfo {year} {2016})}\BibitemShut {NoStop}%
	\bibitem [{\citenamefont {Jacobs}\ \emph {et~al.}(1944)\citenamefont {Jacobs},
		\citenamefont {Roberts},\ and\ \citenamefont
		{MacMillan}}]{jacobs1944dielectric}%
	\BibitemOpen
	\bibfield  {author} {\bibinfo {author} {\bibfnamefont {T.~L.}\ \bibnamefont
			{Jacobs}}, \bibinfo {author} {\bibfnamefont {J.~D.}\ \bibnamefont
			{Roberts}},\ and\ \bibinfo {author} {\bibfnamefont {W.}~\bibnamefont
			{MacMillan}},\ }\href@noop {} {\bibfield  {journal} {\bibinfo  {journal}
			{Journal of the American Chemical Society}\ }\textbf {\bibinfo {volume}
			{66}},\ \bibinfo {pages} {656} (\bibinfo {year} {1944})}\BibitemShut
	{NoStop}%
	\bibitem [{\citenamefont {Gee}\ \emph {et~al.}(1986)\citenamefont {Gee},
		\citenamefont {Shinsaka}, \citenamefont {Dodelet},\ and\ \citenamefont
		{Freeman}}]{gee1986dielectric}%
	\BibitemOpen
	\bibfield  {author} {\bibinfo {author} {\bibfnamefont {N.}~\bibnamefont
			{Gee}}, \bibinfo {author} {\bibfnamefont {K.}~\bibnamefont {Shinsaka}},
		\bibinfo {author} {\bibfnamefont {J.-P.}\ \bibnamefont {Dodelet}},\ and\
		\bibinfo {author} {\bibfnamefont {G.~R.}\ \bibnamefont {Freeman}},\
	}\href@noop {} {\bibfield  {journal} {\bibinfo  {journal} {The Journal of
				Chemical Thermodynamics}\ }\textbf {\bibinfo {volume} {18}},\ \bibinfo
		{pages} {221} (\bibinfo {year} {1986})}\BibitemShut {NoStop}%
	\bibitem [{\citenamefont {Blukis}\ \emph {et~al.}(1963)\citenamefont {Blukis},
		\citenamefont {Kasai},\ and\ \citenamefont {Myers}}]{blukis1963microwave}%
	\BibitemOpen
	\bibfield  {author} {\bibinfo {author} {\bibfnamefont {U.}~\bibnamefont
			{Blukis}}, \bibinfo {author} {\bibfnamefont {P.~H.}\ \bibnamefont {Kasai}},\
		and\ \bibinfo {author} {\bibfnamefont {R.~J.}\ \bibnamefont {Myers}},\
	}\href@noop {} {\bibfield  {journal} {\bibinfo  {journal} {The Journal of
				Chemical Physics}\ }\textbf {\bibinfo {volume} {38}},\ \bibinfo {pages}
		{2753} (\bibinfo {year} {1963})}\BibitemShut {NoStop}%
	\bibitem [{\citenamefont {Eltringham}\ and\ \citenamefont
		{Catchpole}(2007)}]{eltringham2007relative}%
	\BibitemOpen
	\bibfield  {author} {\bibinfo {author} {\bibfnamefont {W.}~\bibnamefont
			{Eltringham}}\ and\ \bibinfo {author} {\bibfnamefont {O.~J.}\ \bibnamefont
			{Catchpole}},\ }\href@noop {} {\bibfield  {journal} {\bibinfo  {journal}
			{Journal of Chemical \& Engineering Data}\ }\textbf {\bibinfo {volume}
			{52}},\ \bibinfo {pages} {363} (\bibinfo {year} {2007})}\BibitemShut
	{NoStop}%
	\bibitem [{\citenamefont {Engerholm}\ \emph {et~al.}(1969)\citenamefont
		{Engerholm}, \citenamefont {Luntz}, \citenamefont {Gwinn},\ and\
		\citenamefont {Harris}}]{engerholm1969ring}%
	\BibitemOpen
	\bibfield  {author} {\bibinfo {author} {\bibfnamefont {G.~G.}\ \bibnamefont
			{Engerholm}}, \bibinfo {author} {\bibfnamefont {A.~C.}\ \bibnamefont
			{Luntz}}, \bibinfo {author} {\bibfnamefont {W.}~\bibnamefont {Gwinn}},\ and\
		\bibinfo {author} {\bibfnamefont {D.~O.}\ \bibnamefont {Harris}},\
	}\href@noop {} {\bibfield  {journal} {\bibinfo  {journal} {The Journal of
				Chemical Physics}\ }\textbf {\bibinfo {volume} {50}},\ \bibinfo {pages}
		{2446} (\bibinfo {year} {1969})}\BibitemShut {NoStop}%
	\bibitem [{\citenamefont {Arivazhagan}\ \emph {et~al.}(2011)\citenamefont
		{Arivazhagan}, \citenamefont {Parthipan},\ and\ \citenamefont
		{Thenappan}}]{arivazhagan2011dielectric}%
	\BibitemOpen
	\bibfield  {author} {\bibinfo {author} {\bibfnamefont {G.}~\bibnamefont
			{Arivazhagan}}, \bibinfo {author} {\bibfnamefont {G.}~\bibnamefont
			{Parthipan}},\ and\ \bibinfo {author} {\bibfnamefont {T.}~\bibnamefont
			{Thenappan}},\ }\href@noop {} {\bibfield  {journal} {\bibinfo  {journal}
			{Physics and Chemistry of Liquids}\ }\textbf {\bibinfo {volume} {49}},\
		\bibinfo {pages} {414} (\bibinfo {year} {2011})}\BibitemShut {NoStop}%
	\bibitem [{\citenamefont {Nain}(2006)}]{nain2006densities}%
	\BibitemOpen
	\bibfield  {author} {\bibinfo {author} {\bibfnamefont {A.~K.}\ \bibnamefont
			{Nain}},\ }\href@noop {} {\bibfield  {journal} {\bibinfo  {journal} {Journal
				of solution chemistry}\ }\textbf {\bibinfo {volume} {35}},\ \bibinfo {pages}
		{1417} (\bibinfo {year} {2006})}\BibitemShut {NoStop}%
	\bibitem [{\citenamefont {Gadhi}\ \emph {et~al.}(1995)\citenamefont {Gadhi},
		\citenamefont {Lahrouni}, \citenamefont {Legrand},\ and\ \citenamefont
		{Demaison}}]{gadhi1995moment}%
	\BibitemOpen
	\bibfield  {author} {\bibinfo {author} {\bibfnamefont {J.}~\bibnamefont
			{Gadhi}}, \bibinfo {author} {\bibfnamefont {A.}~\bibnamefont {Lahrouni}},
		\bibinfo {author} {\bibfnamefont {J.}~\bibnamefont {Legrand}},\ and\ \bibinfo
		{author} {\bibfnamefont {J.}~\bibnamefont {Demaison}},\ }\href@noop {}
	{\bibfield  {journal} {\bibinfo  {journal} {Journal de chimie physique}\
		}\textbf {\bibinfo {volume} {92}},\ \bibinfo {pages} {1984} (\bibinfo {year}
		{1995})}\BibitemShut {NoStop}%
	\bibitem [{\citenamefont {Stoppa}\ \emph {et~al.}(2015)\citenamefont {Stoppa},
		\citenamefont {Nazet}, \citenamefont {Buchner}, \citenamefont {Thoman},\ and\
		\citenamefont {Walther}}]{stoppa2015dielectric}%
	\BibitemOpen
	\bibfield  {author} {\bibinfo {author} {\bibfnamefont {A.}~\bibnamefont
			{Stoppa}}, \bibinfo {author} {\bibfnamefont {A.}~\bibnamefont {Nazet}},
		\bibinfo {author} {\bibfnamefont {R.}~\bibnamefont {Buchner}}, \bibinfo
		{author} {\bibfnamefont {A.}~\bibnamefont {Thoman}},\ and\ \bibinfo {author}
		{\bibfnamefont {M.}~\bibnamefont {Walther}},\ }\href@noop {} {\bibfield
		{journal} {\bibinfo  {journal} {Journal of Molecular Liquids}\ }\textbf
		{\bibinfo {volume} {212}},\ \bibinfo {pages} {963} (\bibinfo {year}
		{2015})}\BibitemShut {NoStop}%
	\bibitem [{\citenamefont {Schmid}\ \emph {et~al.}(2011)\citenamefont {Schmid},
		\citenamefont {Eichenberger}, \citenamefont {Choutko}, \citenamefont
		{Riniker}, \citenamefont {Winger}, \citenamefont {Mark},\ and\ \citenamefont
		{Van~Gunsteren}}]{schmid2011definition}%
	\BibitemOpen
	\bibfield  {author} {\bibinfo {author} {\bibfnamefont {N.}~\bibnamefont
			{Schmid}}, \bibinfo {author} {\bibfnamefont {A.~P.}\ \bibnamefont
			{Eichenberger}}, \bibinfo {author} {\bibfnamefont {A.}~\bibnamefont
			{Choutko}}, \bibinfo {author} {\bibfnamefont {S.}~\bibnamefont {Riniker}},
		\bibinfo {author} {\bibfnamefont {M.}~\bibnamefont {Winger}}, \bibinfo
		{author} {\bibfnamefont {A.~E.}\ \bibnamefont {Mark}},\ and\ \bibinfo
		{author} {\bibfnamefont {W.~F.}\ \bibnamefont {Van~Gunsteren}},\ }\href@noop
	{} {\bibfield  {journal} {\bibinfo  {journal} {European biophysics journal}\
		}\textbf {\bibinfo {volume} {40}},\ \bibinfo {pages} {843} (\bibinfo {year}
		{2011})}\BibitemShut {NoStop}%
	\bibitem [{\citenamefont {Chen}\ \emph {et~al.}(2020)\citenamefont {Chen},
		\citenamefont {Korblova}, \citenamefont {Dong}, \citenamefont {Wei},
		\citenamefont {Shao}, \citenamefont {Radzihovsky}, \citenamefont {Glaser},
		\citenamefont {Maclennan}, \citenamefont {Bedrov}, \citenamefont {Walba}
		\emph {et~al.}}]{chen2020first}%
	\BibitemOpen
	\bibfield  {author} {\bibinfo {author} {\bibfnamefont {X.}~\bibnamefont
			{Chen}}, \bibinfo {author} {\bibfnamefont {E.}~\bibnamefont {Korblova}},
		\bibinfo {author} {\bibfnamefont {D.}~\bibnamefont {Dong}}, \bibinfo {author}
		{\bibfnamefont {X.}~\bibnamefont {Wei}}, \bibinfo {author} {\bibfnamefont
			{R.}~\bibnamefont {Shao}}, \bibinfo {author} {\bibfnamefont {L.}~\bibnamefont
			{Radzihovsky}}, \bibinfo {author} {\bibfnamefont {M.~A.}\ \bibnamefont
			{Glaser}}, \bibinfo {author} {\bibfnamefont {J.~E.}\ \bibnamefont
			{Maclennan}}, \bibinfo {author} {\bibfnamefont {D.}~\bibnamefont {Bedrov}},
		\bibinfo {author} {\bibfnamefont {D.~M.}\ \bibnamefont {Walba}}, \emph
		{et~al.},\ }\href@noop {} {\bibfield  {journal} {\bibinfo  {journal}
			{Proceedings of the National Academy of Sciences}\ }\textbf {\bibinfo
			{volume} {117}},\ \bibinfo {pages} {14021} (\bibinfo {year}
		{2020})}\BibitemShut {NoStop}%
	\bibitem [{\citenamefont {Simeral}\ and\ \citenamefont
		{Amey}(1970)}]{simeral1970dielectric}%
	\BibitemOpen
	\bibfield  {author} {\bibinfo {author} {\bibfnamefont {L.}~\bibnamefont
			{Simeral}}\ and\ \bibinfo {author} {\bibfnamefont {R.~L.}\ \bibnamefont
			{Amey}},\ }\href@noop {} {\bibfield  {journal} {\bibinfo  {journal} {The
				Journal of Physical Chemistry}\ }\textbf {\bibinfo {volume} {74}},\ \bibinfo
		{pages} {1443} (\bibinfo {year} {1970})}\BibitemShut {NoStop}%
	\bibitem [{\citenamefont {Sengwa}\ \emph {et~al.}(2017)\citenamefont {Sengwa},
		\citenamefont {Choudhary},\ and\ \citenamefont
		{Dhatarwal}}]{sengwa2017characterization}%
	\BibitemOpen
	\bibfield  {author} {\bibinfo {author} {\bibfnamefont {R.}~\bibnamefont
			{Sengwa}}, \bibinfo {author} {\bibfnamefont {S.}~\bibnamefont {Choudhary}},\
		and\ \bibinfo {author} {\bibfnamefont {P.}~\bibnamefont {Dhatarwal}},\
	}\href@noop {} {\bibfield  {journal} {\bibinfo  {journal} {Journal of
				Molecular Liquids}\ }\textbf {\bibinfo {volume} {225}},\ \bibinfo {pages}
		{42} (\bibinfo {year} {2017})}\BibitemShut {NoStop}%
	\bibitem [{\citenamefont {Koperwas}\ and\ \citenamefont
		{Paluch}(2022)}]{koperwas2022computational}%
	\BibitemOpen
	\bibfield  {author} {\bibinfo {author} {\bibfnamefont {K.}~\bibnamefont
			{Koperwas}}\ and\ \bibinfo {author} {\bibfnamefont {M.}~\bibnamefont
			{Paluch}},\ }\href@noop {} {\bibfield  {journal} {\bibinfo  {journal}
			{Physical Review Letters}\ }\textbf {\bibinfo {volume} {129}},\ \bibinfo
		{pages} {025501} (\bibinfo {year} {2022})}\BibitemShut {NoStop}%
	\bibitem [{\citenamefont {Pabst}\ \emph {et~al.}(2021)\citenamefont {Pabst},
		\citenamefont {Gabriel}, \citenamefont {B{\"o}hmer}, \citenamefont {Weigl},
		\citenamefont {Helbling}, \citenamefont {Richter}, \citenamefont {Zourchang},
		\citenamefont {Walther},\ and\ \citenamefont
		{Blochowicz}}]{pabst2021generic}%
	\BibitemOpen
	\bibfield  {author} {\bibinfo {author} {\bibfnamefont {F.}~\bibnamefont
			{Pabst}}, \bibinfo {author} {\bibfnamefont {J.~P.}\ \bibnamefont {Gabriel}},
		\bibinfo {author} {\bibfnamefont {T.}~\bibnamefont {B{\"o}hmer}}, \bibinfo
		{author} {\bibfnamefont {P.}~\bibnamefont {Weigl}}, \bibinfo {author}
		{\bibfnamefont {A.}~\bibnamefont {Helbling}}, \bibinfo {author}
		{\bibfnamefont {T.}~\bibnamefont {Richter}}, \bibinfo {author} {\bibfnamefont
			{P.}~\bibnamefont {Zourchang}}, \bibinfo {author} {\bibfnamefont
			{T.}~\bibnamefont {Walther}},\ and\ \bibinfo {author} {\bibfnamefont
			{T.}~\bibnamefont {Blochowicz}},\ }\href@noop {} {\bibfield  {journal}
		{\bibinfo  {journal} {The Journal of Physical Chemistry Letters}\ }\textbf
		{\bibinfo {volume} {12}},\ \bibinfo {pages} {3685} (\bibinfo {year}
		{2021})}\BibitemShut {NoStop}%
	\bibitem [{\citenamefont {H{\'e}not}\ \emph {et~al.}(2023)\citenamefont
		{H{\'e}not}, \citenamefont {D{\'e}jardin},\ and\ \citenamefont
		{Ladieu}}]{henot2023orientational}%
	\BibitemOpen
	\bibfield  {author} {\bibinfo {author} {\bibfnamefont {M.}~\bibnamefont
			{H{\'e}not}}, \bibinfo {author} {\bibfnamefont {P.-M.}\ \bibnamefont
			{D{\'e}jardin}},\ and\ \bibinfo {author} {\bibfnamefont {F.}~\bibnamefont
			{Ladieu}},\ }\href@noop {} {\bibfield  {journal} {\bibinfo  {journal}
			{Physical Chemistry Chemical Physics}\ }\textbf {\bibinfo {volume} {25}},\
		\bibinfo {pages} {29233} (\bibinfo {year} {2023})}\BibitemShut {NoStop}%
	\bibitem [{\citenamefont {Matyushov}\ and\ \citenamefont
		{Richert}(2023)}]{matyushov2023single}%
	\BibitemOpen
	\bibfield  {author} {\bibinfo {author} {\bibfnamefont {D.~V.}\ \bibnamefont
			{Matyushov}}\ and\ \bibinfo {author} {\bibfnamefont {R.}~\bibnamefont
			{Richert}},\ }\href@noop {} {\bibfield  {journal} {\bibinfo  {journal} {The
				journal of physical chemistry letters}\ }\textbf {\bibinfo {volume} {14}},\
		\bibinfo {pages} {4886} (\bibinfo {year} {2023})}\BibitemShut {NoStop}%
	\bibitem [{\citenamefont {B{\"o}hmer}\ \emph {et~al.}(2025)\citenamefont
		{B{\"o}hmer}, \citenamefont {Pabst}, \citenamefont {Gabriel}, \citenamefont
		{Zei{\ss}ler},\ and\ \citenamefont {Blochowicz}}]{bohmer2024spectral}%
	\BibitemOpen
	\bibfield  {author} {\bibinfo {author} {\bibfnamefont {T.}~\bibnamefont
			{B{\"o}hmer}}, \bibinfo {author} {\bibfnamefont {F.}~\bibnamefont {Pabst}},
		\bibinfo {author} {\bibfnamefont {J.~P.}\ \bibnamefont {Gabriel}}, \bibinfo
		{author} {\bibfnamefont {R.}~\bibnamefont {Zei{\ss}ler}},\ and\ \bibinfo
		{author} {\bibfnamefont {T.}~\bibnamefont {Blochowicz}},\ }\href@noop {}
	{\bibfield  {journal} {\bibinfo  {journal} {The Journal of Chemical Physics}\
		}\textbf {\bibinfo {volume} {162}} (\bibinfo {year} {2025})}\BibitemShut
	{NoStop}%
	\bibitem [{\citenamefont {B{\"o}hmer}\ \emph {et~al.}(2024)\citenamefont
		{B{\"o}hmer}, \citenamefont {Pabst}, \citenamefont {Gabriel},\ and\
		\citenamefont {Blochowicz}}]{bohmer2024dipolar}%
	\BibitemOpen
	\bibfield  {author} {\bibinfo {author} {\bibfnamefont {T.}~\bibnamefont
			{B{\"o}hmer}}, \bibinfo {author} {\bibfnamefont {F.}~\bibnamefont {Pabst}},
		\bibinfo {author} {\bibfnamefont {J.~P.}\ \bibnamefont {Gabriel}},\ and\
		\bibinfo {author} {\bibfnamefont {T.}~\bibnamefont {Blochowicz}},\
	}\href@noop {} {\bibfield  {journal} {\bibinfo  {journal} {Physical Review
				Letters}\ }\textbf {\bibinfo {volume} {132}},\ \bibinfo {pages} {206101}
		(\bibinfo {year} {2024})}\BibitemShut {NoStop}%
	\bibitem [{\citenamefont {H{\'e}not}\ and\ \citenamefont
		{Ladieu}(2025)}]{henot2025emergence}%
	\BibitemOpen
	\bibfield  {author} {\bibinfo {author} {\bibfnamefont {M.}~\bibnamefont
			{H{\'e}not}}\ and\ \bibinfo {author} {\bibfnamefont {F.}~\bibnamefont
			{Ladieu}},\ }\href@noop {} {\bibfield  {journal} {\bibinfo  {journal}
			{Physical Review Letters}\ }\textbf {\bibinfo {volume} {134}},\ \bibinfo
		{pages} {218202} (\bibinfo {year} {2025})}\BibitemShut {NoStop}%
	\bibitem [{\citenamefont {Eastman}\ \emph {et~al.}(2017)\citenamefont
		{Eastman}, \citenamefont {Swails}, \citenamefont {Chodera}, \citenamefont
		{McGibbon}, \citenamefont {Zhao}, \citenamefont {Beauchamp}, \citenamefont
		{Wang}, \citenamefont {Simmonett}, \citenamefont {Harrigan}, \citenamefont
		{Stern}, \citenamefont {Wiewiora}, \citenamefont {Brooks},\ and\
		\citenamefont {Pande}}]{openmm_2017}%
	\BibitemOpen
	\bibfield  {author} {\bibinfo {author} {\bibfnamefont {P.}~\bibnamefont
			{Eastman}}, \bibinfo {author} {\bibfnamefont {J.}~\bibnamefont {Swails}},
		\bibinfo {author} {\bibfnamefont {J.}~\bibnamefont {Chodera}}, \bibinfo
		{author} {\bibfnamefont {R.}~\bibnamefont {McGibbon}}, \bibinfo {author}
		{\bibfnamefont {Y.}~\bibnamefont {Zhao}}, \bibinfo {author} {\bibfnamefont
			{K.}~\bibnamefont {Beauchamp}}, \bibinfo {author} {\bibfnamefont {L.-P.}\
			\bibnamefont {Wang}}, \bibinfo {author} {\bibfnamefont {A.}~\bibnamefont
			{Simmonett}}, \bibinfo {author} {\bibfnamefont {M.}~\bibnamefont {Harrigan}},
		\bibinfo {author} {\bibfnamefont {C.}~\bibnamefont {Stern}}, \bibinfo
		{author} {\bibfnamefont {R.}~\bibnamefont {Wiewiora}}, \bibinfo {author}
		{\bibfnamefont {B.}~\bibnamefont {Brooks}},\ and\ \bibinfo {author}
		{\bibfnamefont {V.}~\bibnamefont {Pande}},\ }\href
	{https://doi.org/10.1371/journal.pcbi.1005659} {\bibfield  {journal}
		{\bibinfo  {journal} {PLOS Computational Biology}\ }\textbf {\bibinfo
			{volume} {13}},\ \bibinfo {pages} {1} (\bibinfo {year} {2017})}\BibitemShut
	{NoStop}%
	\bibitem [{\citenamefont {Essmann}\ \emph {et~al.}(1995)\citenamefont
		{Essmann}, \citenamefont {Perera}, \citenamefont {Berkowitz}, \citenamefont
		{Darden}, \citenamefont {Lee},\ and\ \citenamefont
		{Pedersen}}]{essmann1995smooth}%
	\BibitemOpen
	\bibfield  {author} {\bibinfo {author} {\bibfnamefont {U.}~\bibnamefont
			{Essmann}}, \bibinfo {author} {\bibfnamefont {L.}~\bibnamefont {Perera}},
		\bibinfo {author} {\bibfnamefont {M.~L.}\ \bibnamefont {Berkowitz}}, \bibinfo
		{author} {\bibfnamefont {T.}~\bibnamefont {Darden}}, \bibinfo {author}
		{\bibfnamefont {H.}~\bibnamefont {Lee}},\ and\ \bibinfo {author}
		{\bibfnamefont {L.~G.}\ \bibnamefont {Pedersen}},\ }\href@noop {} {\bibfield
		{journal} {\bibinfo  {journal} {The Journal of chemical physics}\ }\textbf
		{\bibinfo {volume} {103}},\ \bibinfo {pages} {8577} (\bibinfo {year}
		{1995})}\BibitemShut {NoStop}%
	\bibitem [{die(2026)}]{dielcav}%
	\BibitemOpen
	\href@noop {} {\bibinfo {title} {Dielcav python package}} (\bibinfo {year}
	{2026}),\ \bibinfo {note}
	{\url{https://github.com/mhenot/dielcav}}\BibitemShut {NoStop}%
	\bibitem [{\citenamefont {Metropolis}\ and\ \citenamefont
		{Ulam}(1949)}]{metropolis1949monte}%
	\BibitemOpen
	\bibfield  {author} {\bibinfo {author} {\bibfnamefont {N.}~\bibnamefont
			{Metropolis}}\ and\ \bibinfo {author} {\bibfnamefont {S.}~\bibnamefont
			{Ulam}},\ }\href@noop {} {\bibfield  {journal} {\bibinfo  {journal} {Journal
				of the American statistical association}\ }\textbf {\bibinfo {volume} {44}},\
		\bibinfo {pages} {335} (\bibinfo {year} {1949})}\BibitemShut {NoStop}%
\end{thebibliography}


\onecolumngrid

\vspace{12pt}
\noindent\hrulefill \hspace{24pt} {\bf Appendix} \hspace{24pt} \hrulefill
\vspace{12pt}

\twocolumngrid

\renewcommand\thefigure{A\arabic{figure}}   
\renewcommand\thetable{A\arabic{table}}    

\renewcommand{\theequation}{A\arabic{equation}}
\setcounter{equation}{0}
\setcounter{figure}{0}

\textbf{A. MD simulation details.} Molecular dynamic simulations were performed using OpenMM~\cite{openmm_2017} in a cubic cell with periodic boundary conditions, in the NVT conditions with a 2~fs time step and a Nosé-Hoover thermostat. Electrostatic interactions were computed with the particle mesh Ewald method~\cite{essmann1995smooth}. The force fields for model diethilether (Et$_2$O-m), dimethilether (Me$_2$O-m), tetrahydrofuran (THF-m) and acetonitrile (MeCN-m) were adapted from GROMOS96 54A7 united atoms~\cite{schmid2011definition} for each molecule in which all partial charges were set to zero, except for the one shown in Fig.~\ref{fig1}a (one negative charge $q$ in red and positive charges in blue insuring neutrality). The diehedral parameters for Et$_2$O-m were also modified to significantly increase the stability of the trans-trans conformer. The molecular volumes $v$ were set to $v=145$~\AA~for Et$_2$O-m, $v=83$~\AA~for Me$_2$O-m, $v=133$~\AA~for THF-m and $v=102$~\AA~for MeCN-m. The number of simulated molecules was $N=2\times 10^4$ for Et$_2$O-m and THF-m and $N=4\times 10^4$ for Me$_2$O-m and MeCN-m. 
Each system at each temperature was equilibrated during a time between 6 and 24 ns corresponding to a at least a few hundreds reorientational time of the dipole and a simulation run was performed for the same amount of time. The output trajectories were analyzed with the virtual cavity method~\cite{henot2025computing} using the dielcav package~\cite{dielcav} and taking the center of mass of each molecule as the cavity center. When the simulation box is large enough as it is the case here, this method gives a direct access to the infinite-system Kirkwood correlation factor $g_\mathrm{K}$ rather than the finite-system one $G_\mathrm{K}=3\epsilon/(2\epsilon+1)g_\mathrm{K}$ which is affected by the use of tin-foil electrostatic boundary conditions~\cite{zhang2016JPCL}.

\textbf{B. Comparison with experimental data.} Experimental data for gas phase dipole moment $\mu_\mathrm{g}$ and the temperature dependence of the dielectric permittivity $\varepsilon(T)$ and density $\rho(T)$ were taken from the literature for each system as detailed in table~\ref{table_A1}. We used the Lorentz-Lorentz equation $(n(T)^2-1)/(n(T)^2+2)\propto \rho(T)$ and the refractive index measured at a reference temperature to calculate the temperature dependent refractive index $n(T)$. In order to ensure consistency with the non-polarizable simulations, the effective dipole moment $\mu(T) = \mu_\mathrm{g} (n(T)^2+2)^2/9$ was used to compute the dimensionless parameter $\lambda_\mathrm{eff}(T) = \mu(T)^2M\mathcal{N}_\mathrm{A}/(3k_\mathrm{K}T\varepsilon_0\rho(T))$. Finally, the Kirkwood correlation factor was obtained from the Kirkwood-Fröhlich equation assuming that $\varepsilon_\infty=n^2$~\cite{kirkwood1939dielectric, frohlich1948general}:
\begin{align}
 g_\mathrm{K}(T) = \frac{(\varepsilon(T) - n(T)^2)(2\varepsilon(T)+n(T)^2)}{3\lambda_\mathrm{eff}(T)\varepsilon(T)}
\end{align}
The results are shown as star markers in Fig.~\ref{fig1}c and as black markers in Fig.~\ref{figS1}.

\begin{table}
	\caption{\label{table_A1} Details on the experimental data. $^a$ Measured under pressure (1.3-33 MPa).}
	\begin{ruledtabular}
		\begin{tabular}{ccc}
			System & $\mu_\mathrm{g}$ (D) & $T$ range for $\varepsilon(T)$ and $\rho(T)$  \\
			\hline
			Et$_2$O   & 1.18~\cite{jacobs1944dielectric} & 171-296 K \cite{gee1986dielectric}  \\
			Me$_2$O   & 1.30~\cite{blukis1963microwave}  & 181-258 K \cite{gee1986dielectric} and 303-383~K~\cite{eltringham2007relative}$^a$\\
			THF       & 1.75~\cite{engerholm1969ring} & 218-297 K \cite{gee1986dielectric} and 303-318 K \cite{arivazhagan2011dielectric, nain2006densities} \\
			MeCN      & 3.92~\cite{gadhi1995moment} & 268-338 K \cite{stoppa2015dielectric} \\
		\end{tabular}
	\end{ruledtabular}
\end{table}

%%%%%%%%%%%%%%%%%%%%%%%%%%%%%%%%%%%%%%%%
\begin{figure}[htbp]
	\centering
	\includegraphics[width=\linewidth]{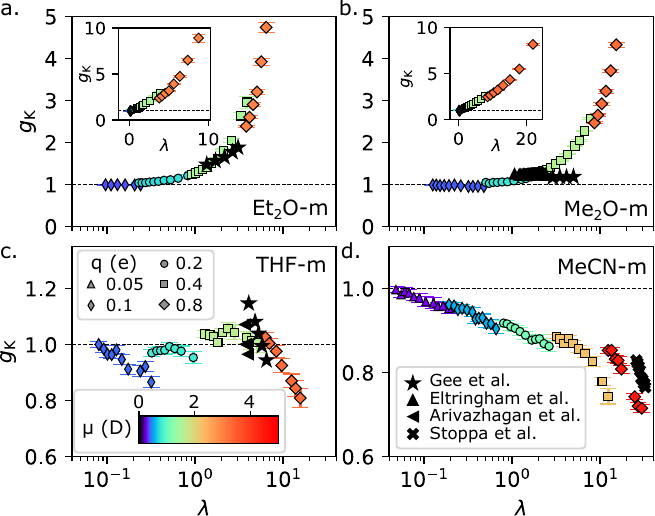}
	\caption{Same plot as in Fig.~\ref{fig1}c for Et$_2$O-m (a), Me$_2$O-m (b), THF-m (c) and MeCN-m (d). Marker shapes indicate the partial charge $q$ (see Fig.~\ref{fig1}a) and colors the dipole moment $\mu$. Experimental data~\cite{jacobs1944dielectric, gee1986dielectric,  blukis1963microwave, eltringham2007relative, engerholm1969ring, arivazhagan2011dielectric, nain2006densities, gadhi1995moment, stoppa2015dielectric} are shown as black markers. Insets in (a) and (b) show the same data on a linear horizontal axis.}
	\label{figS1}
\end{figure}
%%%%%%%%%%%%%%%%%%%%%%%%%%%%%%%%%%%%%%%%

\textbf{C. Dipoles on a lattice.} The simulations of the Hamiltonian given by Eq.~\ref{eq_H_MC} for dipoles on a lattice were performed using the Monte Carlo method with the Metropolis~\cite{metropolis1949monte} algorithm. The system consisted in $N=20~000$ dipoles disposed on a cubic lattice with periodic boundary conditions. The interaction term $J(\theta)$ between neighboring dipoles is not symmetric (\textit{i.e.} $J(\theta_{ij})\neq J(\theta_{ji})$) contrary to the dipolar interaction it intends to mimic. But, because the sum in Eq.~\ref{eq_H_MC} is performed on every dipoles, the final interaction is symmetric. 

\textbf{D. Decomposition of dipolar cross correlations.} For the dipoles on a cubic lattice, the definition of the first two layers used in Fig.~\ref{fig4} is unambiguous: The first layer include the first 6 neighbors appearing in the Hamiltonian and the second layer include the 8 next dipoles corresponding to the corners of the cube whose center is the reference dipole. For the polar liquids simulations, defining these layers is more delicate. We chose to define the limits $r_1$ and $r_2$ of the first and second layers from the first maximum and following minimum of the pair correlation function of the center of mass respectively. In the case of MeCN-m, we could not easily define a meaningful decomposition into spherical layers, likely because of the elongated shape of the molecule.

\end{document}